\documentclass{article} % For LaTeX2e
\usepackage{iclr2025_conference,times}
\iclrfinalcopy
\usepackage{amsmath,amsfonts,bm}

\def\eqref#1{equation~\ref{#1}}
\def\1{\bm{1}}

\DeclareMathAlphabet{\mathsfit}{\encodingdefault}{\sfdefault}{m}{sl}
\SetMathAlphabet{\mathsfit}{bold}{\encodingdefault}{\sfdefault}{bx}{n}

\usepackage{hyperref}
\usepackage{url}
\usepackage{graphicx}
\usepackage{mathrsfs}
\usepackage{verbatim}
\usepackage[utf8]{inputenc}
\usepackage{listings}
\usepackage{geometry}
\lstdefinelanguage{BibTeX}{
  morekeywords={@inproceedings, author, title, booktitle, year, date, note, url},
  sensitive=true,
  morecomment=[l][\itshape\color{gray}]{%},
}
}

\lstdefinelanguage{Reference1}{
  morekeywords={title},
  sensitive=true,
  morecomment=[l][\itshape\color{gray}]{%},
}}

\title{Understanding Model Calibration - A gentle introduction and visual exploration of calibration and the expected calibration error (ECE)}

\author{Maja Pavlovic \\ %\thanks{ Use footnote for providing further information about author (webpage, alternative address)---\emph{not} for acknowledging funding agencies.  Funding acknowledgements go at the end of the paper.
Queen Mary University London\\
\texttt{m.pavlovic@qmul.ac.uk} %\\
}

\begin{document}

\maketitle
\vspace{-0.4cm}
\begin{abstract}
To be considered reliable, a model must be calibrated so that its confidence in each decision closely reflects its true outcome. In this blogpost we'll take a look at the most commonly used definition for calibration and then dive into a frequently used evaluation measure for model calibration. We'll then cover some of the drawbacks of this measure and how these surfaced the need for additional notions of calibration, which require their own new evaluation measures. This post is not intended to be an in-depth dissection of all works on calibration, nor does it focus on how to calibrate models. Instead, it is meant to provide a gentle introduction to the different notions and their evaluation measures as well as to re-highlight some  issues with a measure that is still widely used to evaluate calibration.
\end{abstract}

% \vspace{0.2cm}
\tableofcontents
\clearpage

\section{What is Calibration?}
\textbf{Calibration} makes sure that a model's estimated probabilities match real-world likelihoods. For example, if a weather forecasting model predicts a 70\% chance of rain on several days, then roughly 70\% of those days should actually be rainy for the model to be considered well calibrated \citep{dawid1982well, degroot1983comparison}. This makes model predictions more \textit{reliable} and \textit{trustworthy}, which makes calibration relevant for many applications across various domains.

\begin{figure}[ht]
\begin{center}
    \includegraphics[width=0.75\textwidth]{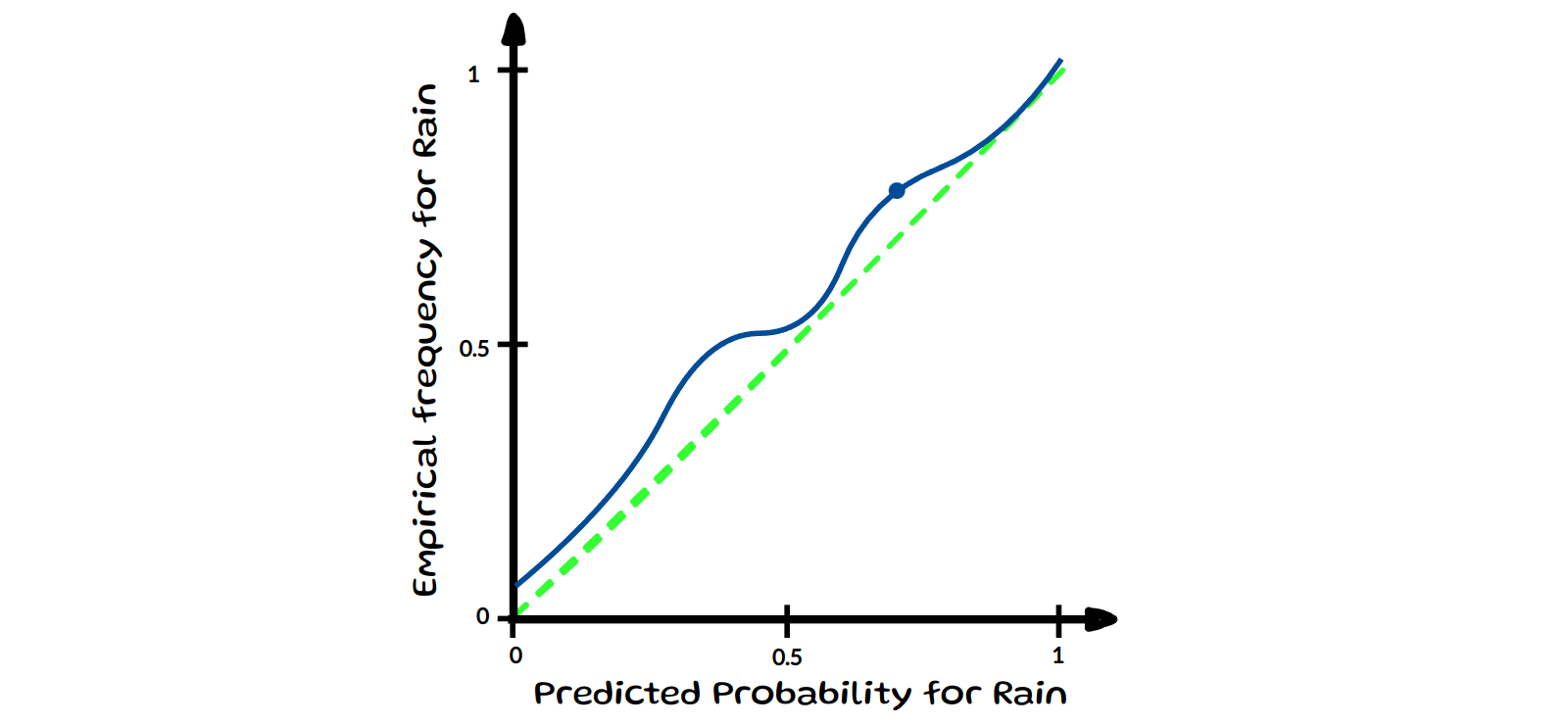}
\end{center}
\caption{Reliability Diagram}
\label{fig:1}
\end{figure}

Now, what \textbf{calibration} means more precisely depends on the specific definition being considered. 
We will have a look at the most common notion in machine learning (ML) formalised in \citep{guo2017calibration} and termed \textbf{\textit{confidence calibration}} in \citep{kull2019beyond}. But first, let's define a bit of formal notation for this blog. 
In this blogpost we consider a classification task with $K$ possible classes, with labels $Y \in \{1, ..., K\}$ and a classification model $\hat{p} : \mathscr{X} \rightarrow \Delta^K$, that takes inputs in $\mathscr{X}$ (e.g. an image or text) and returns a probability vector as its output. $\Delta^K $ refers to the K-simplex, which just means that the elements of the output vector must sum to 1 and that each estimated probability in the vector is between 0 \& 1. These individual probabilities (\textit{or confidences}) indicate how likely 
an input belongs to each of the $K$ classes.

\begin{figure}[ht]
\begin{center}
    \includegraphics[width=0.83\textwidth]{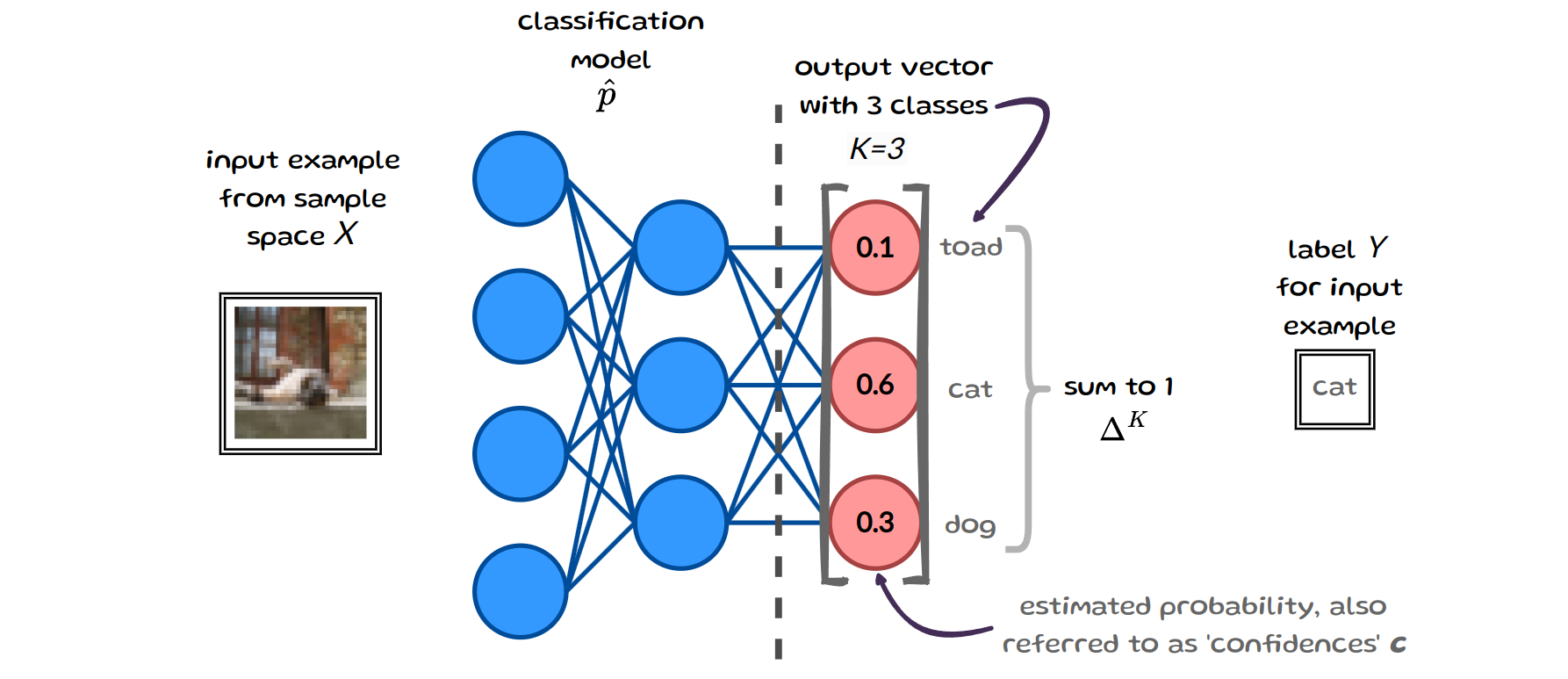}
\end{center}
\caption{Notation - input example sourced from \citep{uma2021learning}}
\label{fig:2}
\end{figure}

\clearpage

\subsection{(Confidence) Calibration}
A model is considered confidence-calibrated if, for all confidences $c,$ the model is correct $c$ proportion of the time:
$$  \mathbb{P} (Y  = \text{arg max}(\hat{p}(X)) \; | \; \text{max}(\hat{p}(X))=c ) = c \;\;\:\:  \forall c \in [0, 1] \; ,$$

where $(X,Y)$ is a datapoint and $\hat{p} : \mathscr{X} \rightarrow \Delta^K$ returns a probability vector as its output. 

This definition of calibration, ensures that the model's final predictions align with their observed accuracy at that 
confidence level \citep{guo2017calibration}. The left chart below visualises the perfectly calibrated outcome (green diagonal line) for all confidences using  a binned reliability diagram \citep{guo2017calibration}. On the right hand side it shows two examples for a specific confidence level across 10 samples.

\begin{figure}[ht]
\begin{center}
    \includegraphics[width=0.9\textwidth]{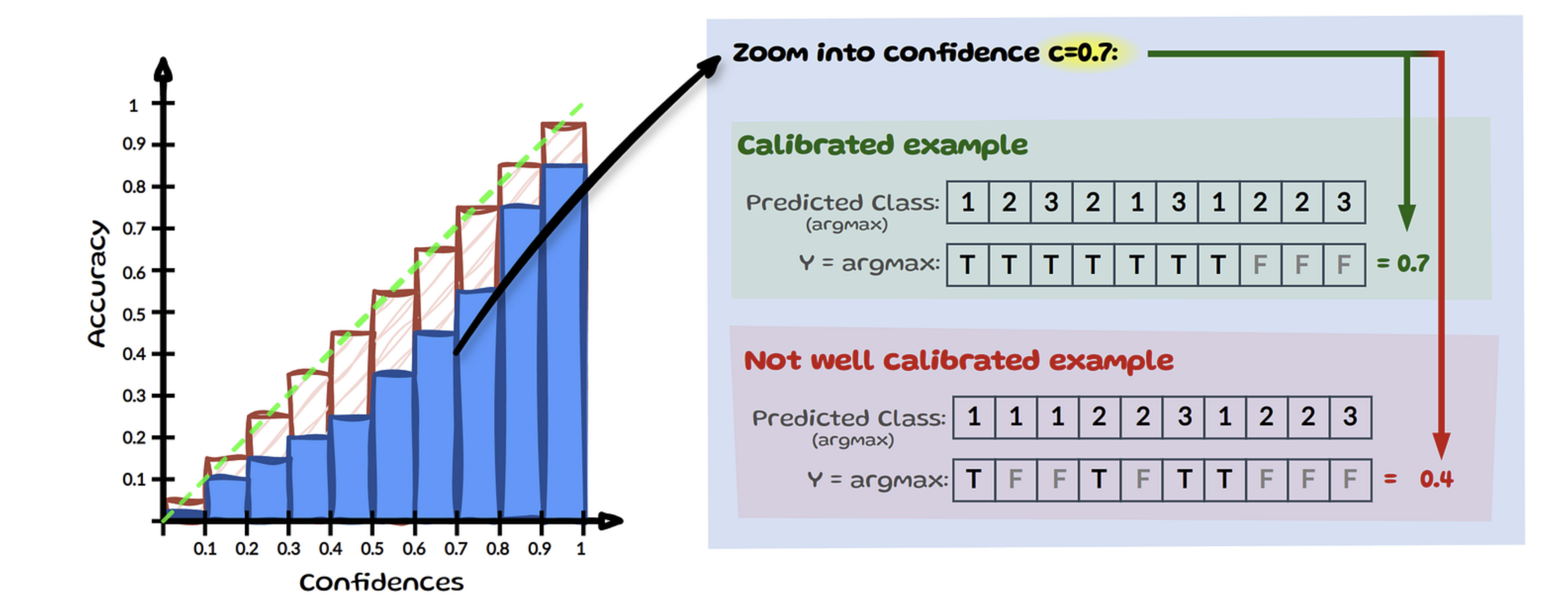}    
\end{center}
\caption{Confidence Calibration}
\label{fig:3}
\end{figure}

For simplification, we assume that we only have 3 classes as in figure 2 and we zoom into confidence $c=0.7$, see image above. Let's assume we have 10 inputs here whose most confident prediction (\textit{max}) equals $0.7$. If the model correctly classifies 7 out of 10 predictions (\textit{true}), it is considered calibrated at confidence level $0.7$. For the model to be fully calibrated this has to hold across all confidence levels from 0 to 1. At the same level $c=0.7$, a model would be considered miscalibrated if it makes only 4 correct predictions.

\section{Evaluating Calibration - Expected Calibration Error (ECE)}

One widely used evaluation measure for confidence calibration is the Expected Calibration Error (ECE) \citep{naeini2015obtaining, guo2017calibration}. ECE measures how well a model's estimated probabilities match the observed probabilities by taking a weighted average over the absolute difference between average accuracy (\textit{acc}) and average confidence (\textit{conf}). The measure involves splitting all $n$ datapoints into M equally spaced bins:
\vspace{-0.07cm}
$$  ECE = \sum_{m=1}^M \frac{\mathopen| B_m \mathclose|}{n} \mathopen| acc(B_m) - conf(B_m) \mathclose| , $$

where $B$ is used for representing "bins" and $m$ for the bin number, while \textit{acc} and \textit{conf} are:
\vspace{-0.07cm}
$${ acc(B_m) = \frac{1}{ \mathopen| B_m \mathclose|} \sum_{i\in B_m} {\displaystyle \1} (\hat{y}_i = y_i ) \;\: \text{\&} \;\: conf(B_m) = \frac{1}{ \mathopen| B_m \mathclose|} \sum_{i\in B_m} \hat{p}(x_i) } $$

$\hat{y}_i$ is the model's predicted class (\textit{arg max}) for sample $i$ and $y_i$ is the true label for sample $i$. ${\displaystyle \1}$ is an indicator function, meaning when the predicted label $\hat{y}_i$ equals the true label $y_i$ it evaluates to 1, otherwise 0.
Let's  look at an example, which will clarify  \textit{acc}, \textit{conf} and the whole binning approach in a visual step-by-step manner.

\subsection{ECE - Visual Step by Step Example}
In the image below, we can see that we have $9$ samples indexed by $i$ with estimated probabilities $\hat{p}(x_i)$ (simplified as $\hat{p}_i$) for class \textbf{cat (C)}, \textbf{dog (D)} or \textbf{toad (T)}. The final column shows the true class ${y}_i$ and the penultimate column contains the predicted class $\hat{y}_i$.

\begin{figure}[ht]
\begin{center}
    \includegraphics[width=0.8\textwidth]{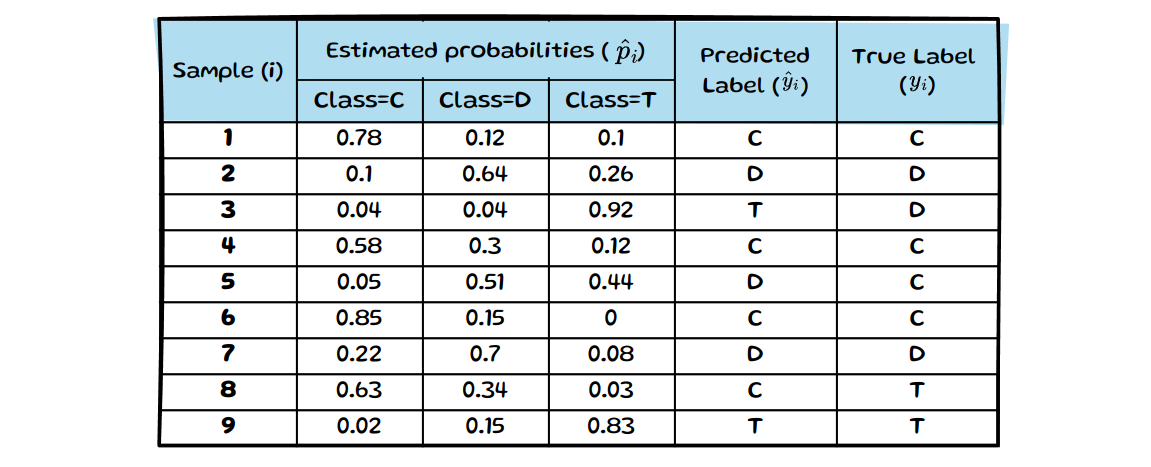}
\end{center}
\caption{ECE - Toy Example}
\label{fig:4}
\end{figure}

Only the maximum probabilities, which determine the predicted label are used in ECE \citep{guo2017calibration}. Therefore, we will only bin samples based on the maximum probability across classes (\textit{see left table in below image}). To keep the example simple we split the data into 5 \textbf{equally spaced} bins $M=5$. If we now look at each sample's maximum estimated probability, we can group it into one of the 5 bins (\textit{see right side of image below}).

\begin{figure}[ht]
\begin{center}
    \includegraphics[width=0.95\textwidth]{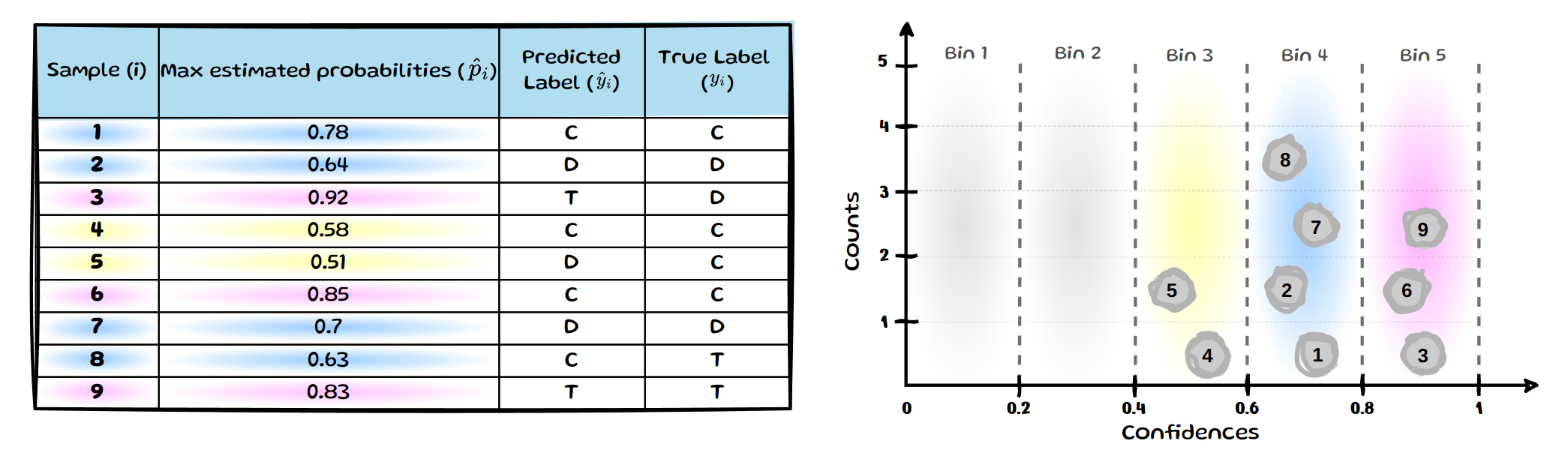}
\end{center}
\caption{Table 2 \& Binning Diagram}
\label{fig:5}
\end{figure}

% \newpage

We still need to determine if the predicted class is correct or not to be able to determine the average accuracy per bin. If the model predicts the class correctly (i.e. $y_i =\hat{y}_i$), the prediction is highlighted in green; incorrect predictions are marked in red:

\newpage

\begin{figure}[ht]
\begin{center}
    \includegraphics[width=0.95\textwidth]{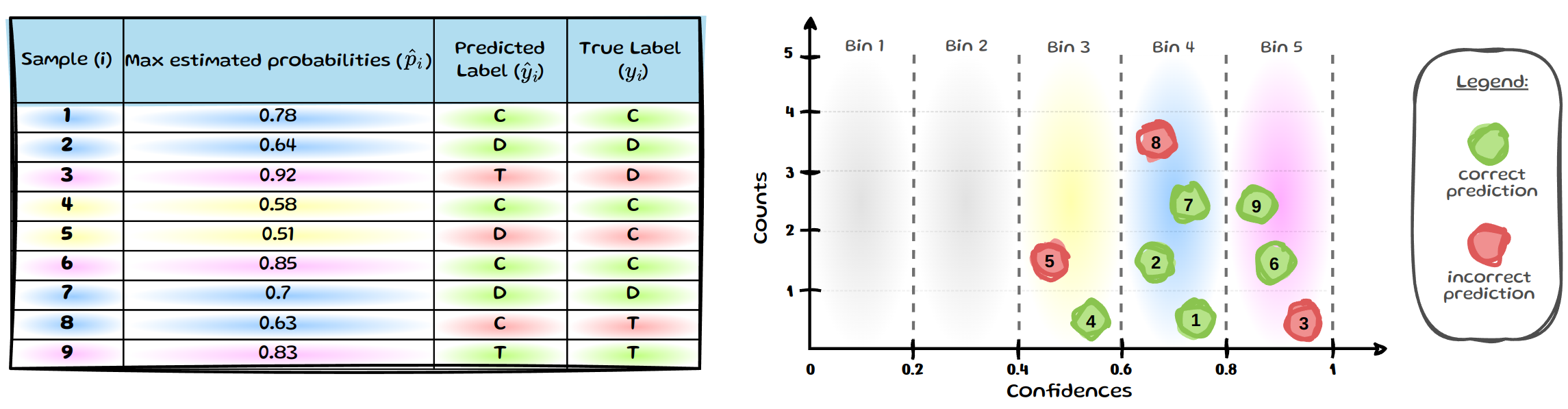}
\end{center}
\caption{Table 3 \& Binning Diagram}
\label{fig:6}
\end{figure}

We now have visualised all the information needed for ECE and will briefly run through how to calculate the values for bin 5 ($B_5$). The other bins then simply follow the same process, see below.

\begin{figure}[ht]
\begin{center}
    \includegraphics[width=1\textwidth]{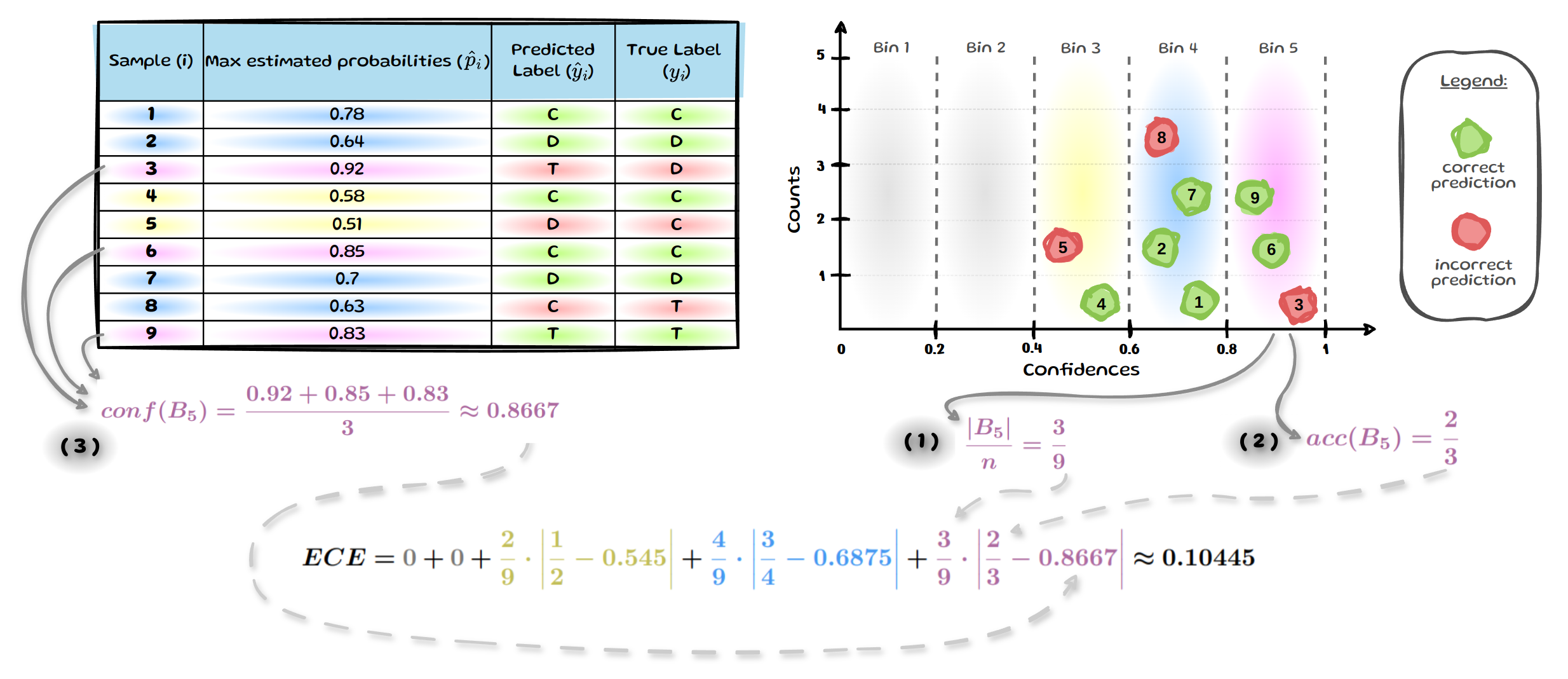}
\end{center}
\caption{Table 4 \& Example for bin 5}
\label{fig:7}
\end{figure}

We can get the empirical probability of a sample falling into $B_5$, by assessing how many out of all $9$ samples fall into $B_5$, see $\mathbf{(\;1\;)}$. We then get the average accuracy for $B_5$, see $\mathbf{(\;2\;)}$ and lastly the average estimated probability for $B_5$, see $\mathbf{(\;3\;)}$. Repeat this for all bins and in our small example of $9$ samples we end up with an ECE of $0.10445$. A perfectly calibrated model would have an ECE of 0.

\subsubsection{Expected Calibration Error Drawbacks}
The images of binning above provide a visual guide of how ECE could result in very different values if we used more bins or perhaps binned the same number of items instead of using equal bin widths. Such and more drawbacks of ECE have been highlighted by several works early on \citep{kumar2018trainable, nixon2019measuring, gupta2020calibration, zhang2020mix, roelofs2022mitigating, vaicenavicius2019evaluating, widmann2019calibration, futami2024informationtheoretic}. However, despite the known weaknesses ECE is still widely used to evaluate confidence calibration in ML \citep{xiong2023can, yuan2024does, collins2023human, si2023prompting, mukhoti2023deep, gao2024spuq}. This motivated this blogpost, with the idea to highlight the most frequently mentioned drawbacks of ECE visually and to provide a simple clarification on the development of different notions of calibration.

\section{Most frequently mentioned Drawbacks of ECE}
\subsection{Pathologies - Low ECE does not equal high accuracy}
A model which minimises ECE, does not necessarily have a high accuracy \citep{kumar2018trainable, kull2019beyond, si2022re}. 
For instance, if a model always predicts the majority class with that class's average prevalence as the probability, it will have an ECE of 0. This is visualised in the image above, where we have a dataset with 10 samples, 7 of those are cat, 2 dog and only one is a toad. Now if the model always predicts cat with on average 0.7 confidence it would have an ECE of 0. 
There are more of such pathologies \citep{nixon2019measuring}. To not only rely on ECE, some researchers use additional measures such as the Brier score or LogLoss alongside ECE \citep{kumar2018trainable, kull2019beyond}.
\begin{figure}[ht]
\begin{center}
    \includegraphics[width=0.85\textwidth]{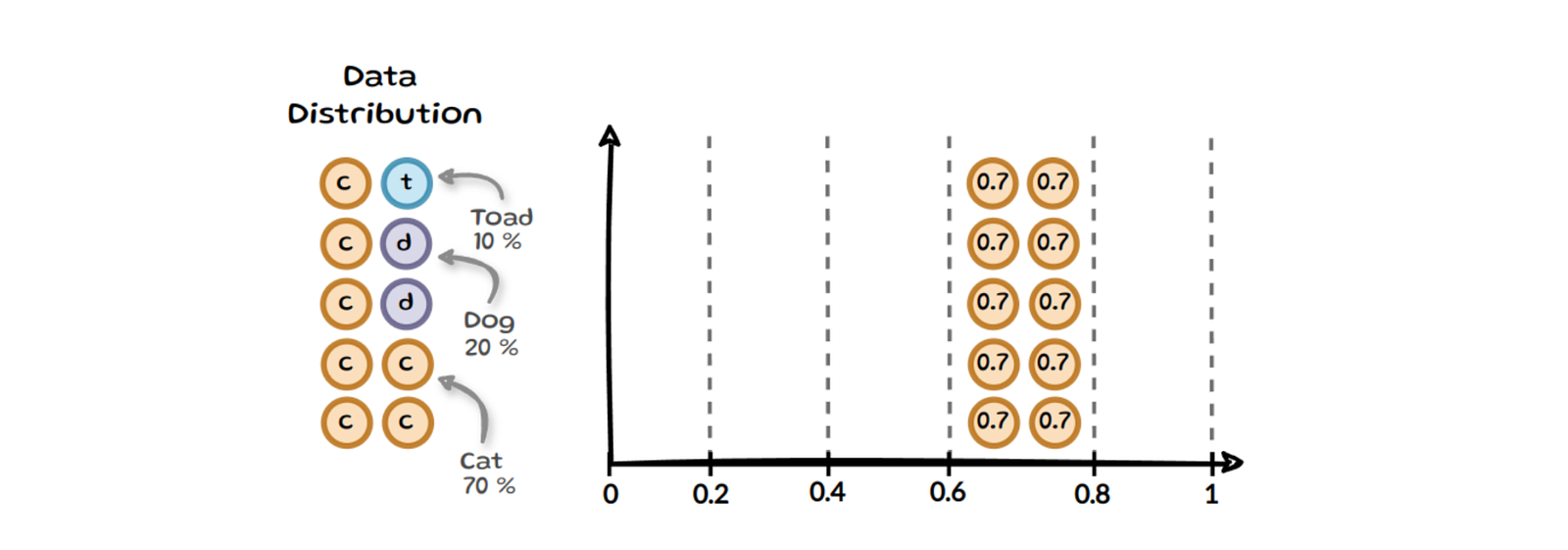}
\end{center}
\caption{Pathologies Example}
\label{fig:8}
\end{figure}

\subsection{Binning Approach}

One of the most frequently mentioned issues with ECE is its sensitivity to the change in binning \citep{kumar2018trainable, nixon2019measuring, gupta2020calibration, zhang2020mix, roelofs2022mitigating, famiglini2023towards}. 
This is sometimes referred to as the \textbf{\textit{Bias-Variance trade-off}} \citep{nixon2019measuring, zhang2020mix}: 
Fewer bins reduce variance but increase bias, while more bins lead to sparsely populated bins increasing variance. 
If we look back to our ECE example with 9 samples and change the bins from 5 to 10 here too, 
we end up with the following:
\begin{figure}[ht]
\begin{center}
    \includegraphics[width=0.95\textwidth]{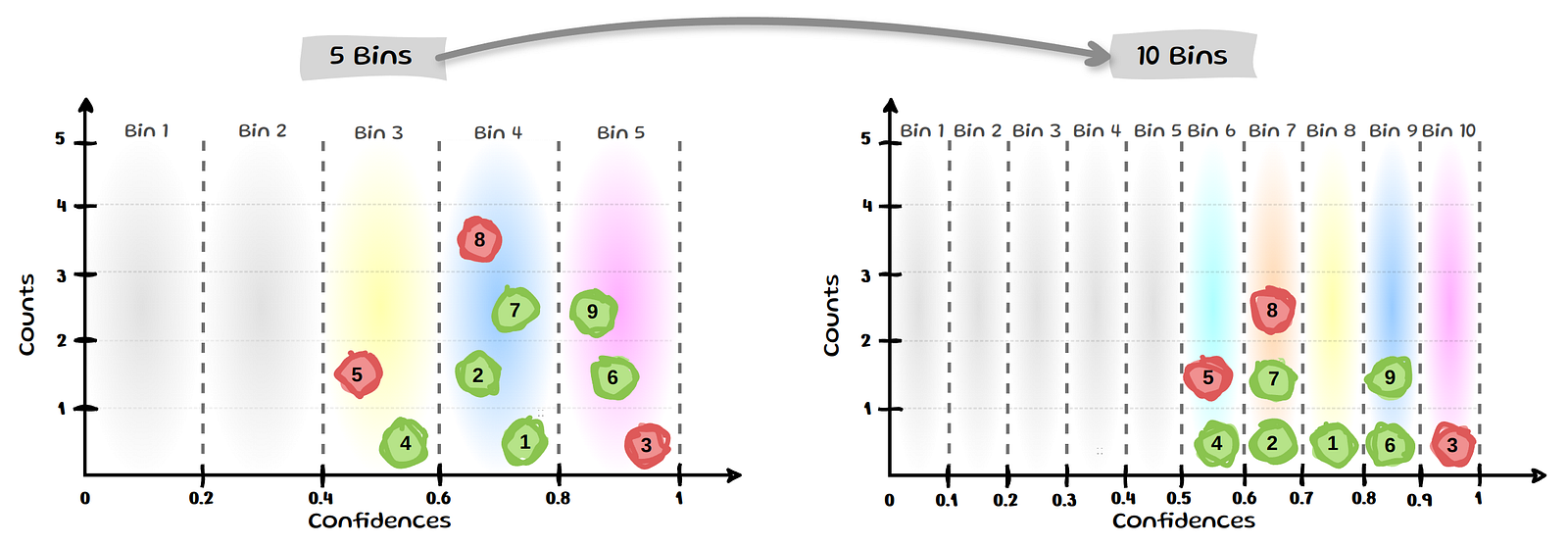}
\end{center}
\caption{More Bins}
\label{fig:9}
\end{figure}

\newpage
\vspace{-1cm}
We can see that bin \textit{8} and \textit{9} each contain only a single sample and also that half the bins now contain 
no samples. The above is only a toy example, however since modern models tend to have higher confidence values 
samples often end up in the last few bins \citep{naeini2015obtaining, zhang2020mix}, which 
means they get all the weight in ECE, while the average error for the empty bins contributes 0 to ECE.

To mitigate these issues of fixed bin widths some authors \citep{nixon2019measuring, roelofs2022mitigating} have proposed a more adaptive binning approach.
\begin{figure}[ht]
\begin{center}
    \includegraphics[width=0.95\textwidth]{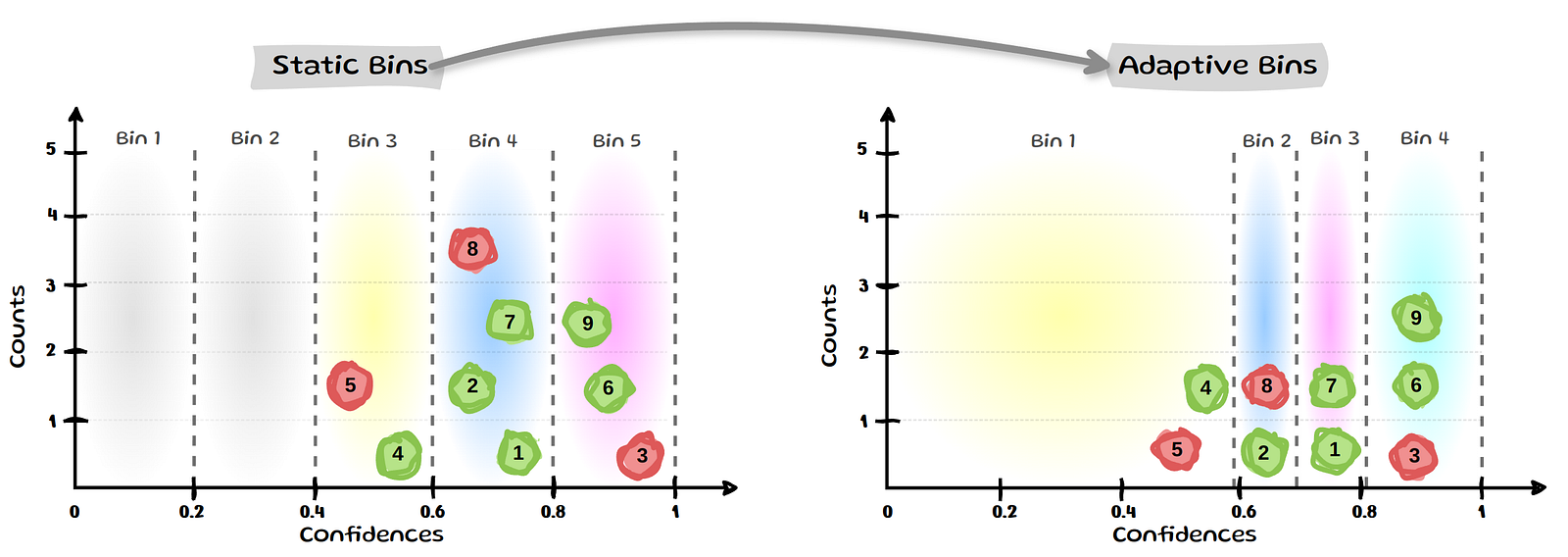}
\end{center}
\vspace{-0.1cm}
\caption{Adaptive Bins}
\label{fig:10}
\end{figure}

Binning-based evaluation with bins containing an equal number of samples are shown to have 
\textit{lower bias} than a fixed binning approach such as ECE \citep{roelofs2022mitigating}. 
This leads \citep{roelofs2022mitigating} to urge against using equal width binning and to
suggest the use of an alternative: ECEsweep, which maximizes the number of equal-mass bins while ensuring 
the calibration function remains monotonic \citep{roelofs2022mitigating}. 
The Adaptive Calibration Error (ACE) and Threshold Adaptive calibration Error (TACE) are two 
other variations of ECE that use flexible binning \citep{nixon2019measuring}.
However, some find it sensitive to the choice of bins and thresholds, leading to inconsistencies in 
ranking different models \citep{ashukha2020pitfalls}. 
Two other approaches aim to eliminate binning altogether: MacroCE does this by averaging over 
instance-level calibration errors of correct and wrong predictions \citep{si2022re} and 
the KDE-based ECE does so by replacing the bins with non-parametric density estimators, 
specifically kernel density estimation (KDE) \citep{zhang2020mix}.

\subsection{Only maximum probabilities considered}
Another frequently mentioned drawback of ECE is that it only considers the maximum estimated probabilities \citep{nixon2019measuring, ashukha2020pitfalls, vaicenavicius2019evaluating, widmann2019calibration, kull2019beyond}. The idea that more than just the maximum confidence should be calibrated, is best illustrated with a simple example \citep{vaicenavicius2019evaluating}:

\newpage
\begin{figure}[ht!]
\begin{center}
    \includegraphics[width=0.73\textwidth]{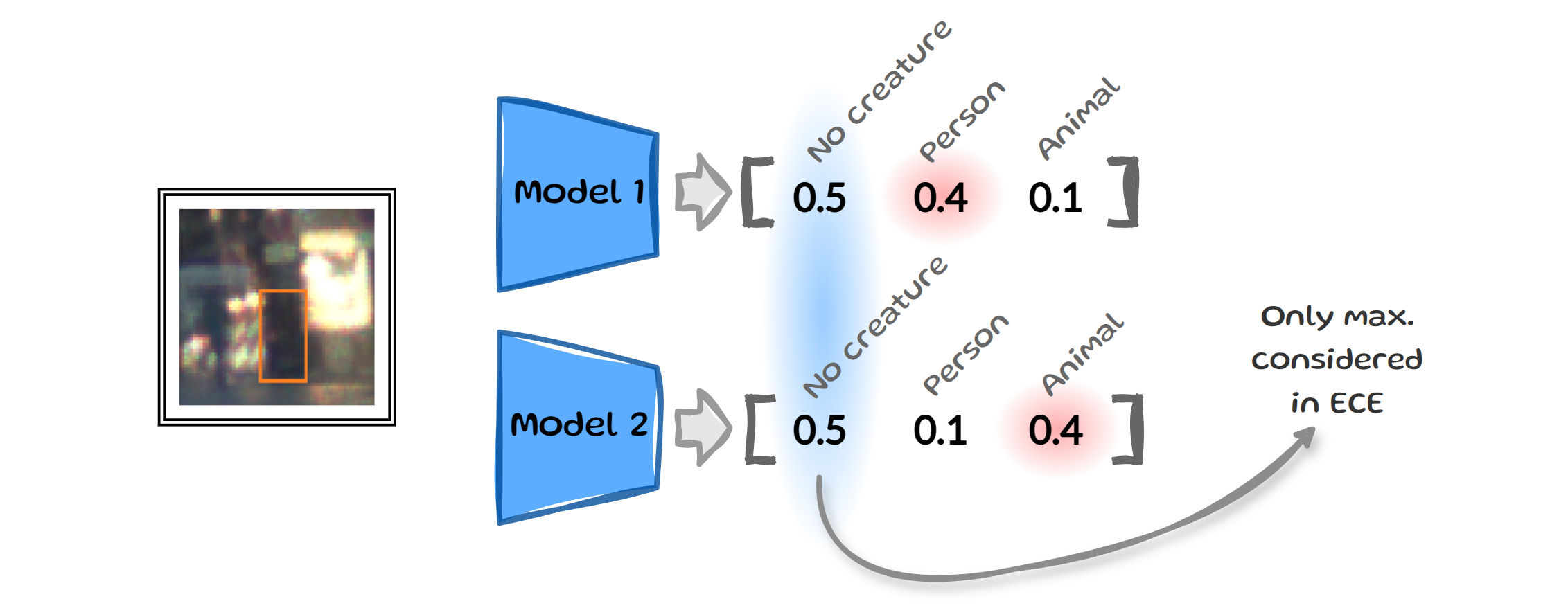}
\end{center}
% \vspace{-0.1cm}
\caption{input example sourced from \citep{schwirten2024ambiguous}}
\label{fig:11}
\end{figure}

% \clearpage

Let's say we trained two different models and now both need to determine if the same input image contains a \textit{person}, an \textit{animal} or \textit{no creature}. The two models output vectors with slightly different estimated probabilities, but both have the same maximum confidence for "\textit{no creature}". Since ECE only looks at these top values it would consider these two outputs to be the same. Yet, when we think of real-world applications we might want our self-driving car to act differently in one situation over the other \citep{vaicenavicius2019evaluating}. This restriction to the maximum confidence prompted various authors \citep{vaicenavicius2019evaluating, kull2019beyond, widmann2019calibration} to reconsider the definition of calibration. 
The existing concept of calibration as "confidence calibration" (coined in \citep{kull2019beyond}) makes a distinction between two additional interpretations of confidence: \textbf{multi-class} and \textbf{class-wise calibration}.

\subsubsection{Multi-class Calibration}

A model is considered multi-class calibrated if, for any prediction vector $q=(q_1,...,q_K) \in \Delta^K $, the class proportions among 
all values of $X$ for which a model outputs the same prediction $\hat{p}(X)=q$ match the values in the prediction vector $q$.

$$  \mathbb{P} (Y  = k \; | \; \hat{p}(X)=q  ) = q_k \;\;\;\:\:\:\:  \forall k \in \{1,...,K\}, \; \forall q \in \Delta^K \, , $$

where $(X,Y)$ is a datapoint and $\hat{p} : \mathscr{X} \rightarrow \Delta^K$ returns a probability vector as its output.

What does this mean in simple terms? Instead of $c$ we now calibrate against a vector $q$, with K classes. Let's look at an example below:
\begin{figure}[ht!]
\begin{center}
    \includegraphics[width=0.89\textwidth]{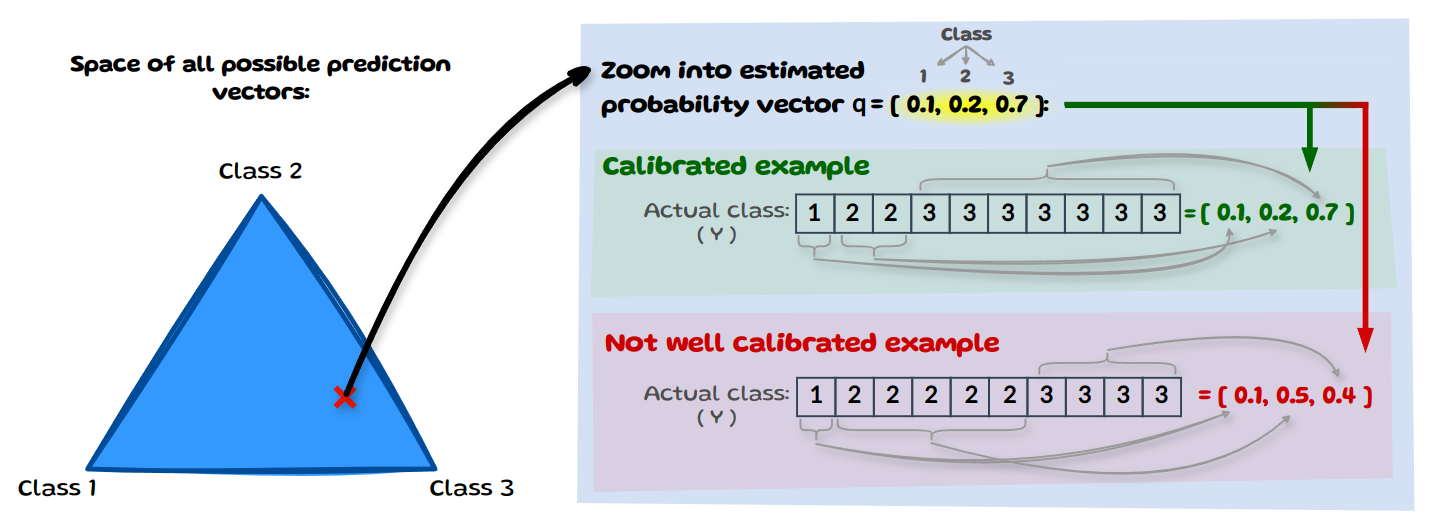}
\end{center}
\caption{Multi-class Calibration}
\label{fig:12}
\end{figure}

\clearpage

On the left we have the space of all possible prediction vectors. Let's zoom into one such vector that our  model predicted and say the model has 10 instances  for which it predicted the vector $q=[0.1,0.2,0.7]$. Now in order for it to be multi-class calibrated, the  distribution of the true (\textit{actual}) class needs to match the prediction vector $q$. The image above shows a calibrated example with $[0.1,0.2,0.7]$ and a not calibrated case with $[0.1,0.5,0.4]$.

\subsubsection{Class-wise Calibration}
A model is considered class-wise calibrated if, 
for each class k, all inputs that share an estimated probability $\hat{p}_k(X)$
align with the true frequency of class k when considered on its own:

$$ \mathbb{P} (Y  = k \; | \; \hat{p}_k(X)= q_k  ) = q_k \;\;\;\;\;\;  \forall k \in \{1,...,K\}  \; , $$

where $(X,Y)$ is a datapoint; $q \in \Delta^K $ and $\hat{p} : \mathscr{X} \rightarrow \Delta^K$ returns a probability vector as its output.

Class-wise calibration is a \textbf{\textit{weaker}} definition than \textbf{multi-class} calibration as it considers each class probability in \textbf{\textit{isolation}} rather than needing the full vector to align. The image below illustrates this by zooming into a probability estimate for class 1 specifically: $q_1=0.1$. Yet again, we assume we have 10 instances for which the model predicted a probability estimate of 0.1 for class 1. We then look at the true class frequency amongst all classes with $q_1=0.1$. If the empirical frequency matches $q_1$ it is calibrated.
\begin{figure}[ht!]
\begin{center}
    \includegraphics[width=0.89\textwidth]{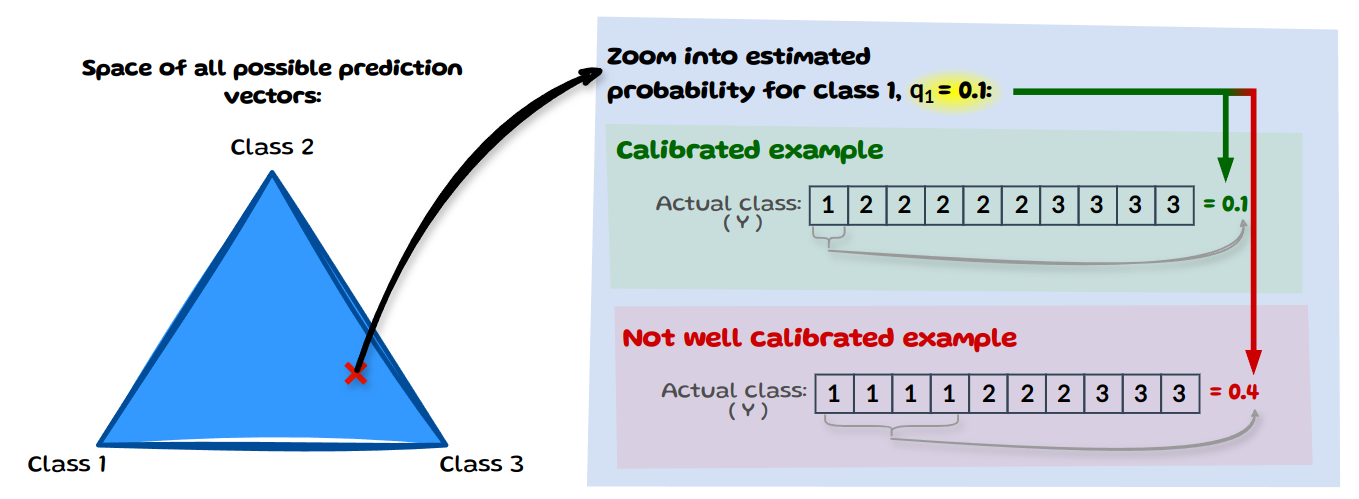}
\end{center}
\caption{Class-wise Calibration}
\label{fig:13}
\end{figure}

To evaluate such different notions of calibration, some updates are made to ECE to calculate a class-wise error. One idea is to calculate the ECE for each class and then take the average \citep{nixon2019measuring, kull2019beyond}. Another idea is to swap the L1-distance used in ECE with the L2 and use several ECE metrics to more effectively assess the overall level of calibration \citep{famiglini2023towards}.
Others, introduce the use of the KS-test for class-wise calibration \citep{gupta2020calibration} and \citep{vaicenavicius2019evaluating} also 
suggest using statistical hypothesis tests instead of ECE based approaches. And other researchers develop a hypothesis test framework (TCal) to detect whether a model is significantly mis-calibrated \citep{donghwan2023tcal} and \citet{sun2024confidenceintervalell2expected} build on this by developing confidence intervals for the L2-ECE.

All the approaches mentioned above \textbf{share a key assumption: ground-truth labels are available}. Within this gold-standard mindset a prediction is either true or false. However, annotators  might unresolvably and justifiably disagree on the real label \citep{aroyo2015truth, uma2021learning}. Let's look at a simple example below:

% \clearpage

\begin{figure}[ht]
\begin{center}
    \includegraphics[width=0.85\textwidth]{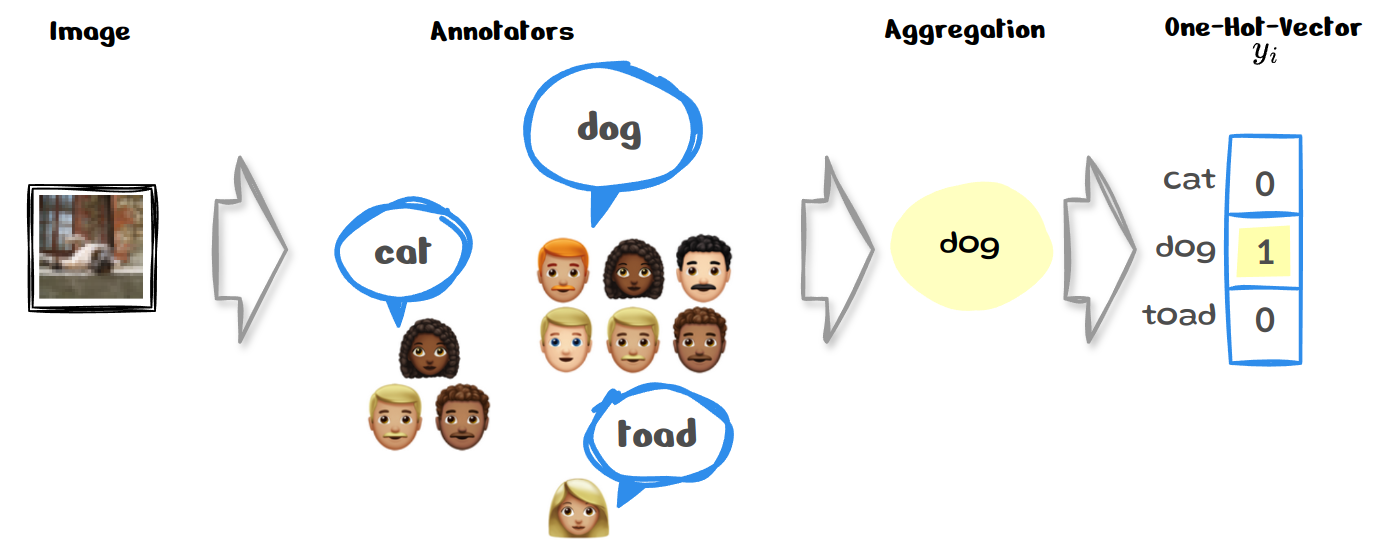}
\end{center}
\caption{One-Hot-Vector}
\label{fig:14}
\end{figure}

% \clearpage

We have the same image as in our entry example and can see that the chosen label differs between annotators. A common approach to resolving such issues in the labelling process is to use some form of aggregation \citep{paun2022statistical, artstein2008inter}. Let's say that in our example the majority vote is selected, so we end up evaluating how well our model is calibrated against such  'ground truth'. One might think, the image is small and pixelated; of course humans will not be certain about their choice. However,  rather than being an exception such disagreements are widespread \citep{aroyo2024dices, sanders2022ambiguous, schwirten2024ambiguous}. So, when there is a lot of human disagreement in a dataset it might not be a good idea to calibrate against an aggregated 'gold' label \citep{baan2022stop}. Instead of gold labels more and more researchers are using soft or smooth labels which are more representative of the human uncertainty \citep{peterson2019cifar, sanders2022ambiguous, collins2023human}, see example below.

\begin{figure}[ht]
\begin{center}
    \includegraphics[width=0.85\textwidth]{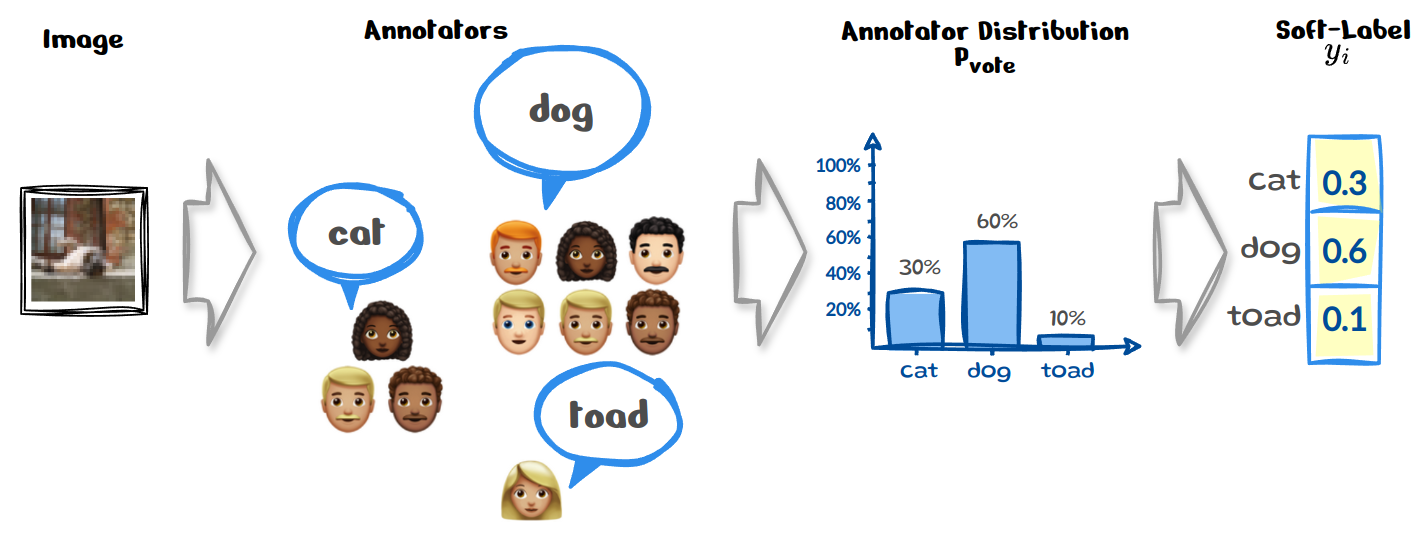}
\end{center}
\caption{Soft-Label}
\label{fig:15}
\end{figure}

In the same example as above, instead of aggregating the annotator votes we could simply use their frequencies to create a distribution $P_{vote}$ over the labels instead, which is then our new $y_i$. This shift towards training models on collective annotator views, rather than relying on a single source-of-truth motivates another definition of calibration: calibrating the model against human uncertainty \citep{baan2022stop}.

\subsubsection{Human Uncertainty Calibration}

A model is considered human-uncertainty calibrated if, for each specific sample $x$, the predicted probability for each class k matches the '\textit{actual}' probability $P_{vote}$ of that class being correct.

$$ \mathbb{P}_{vote} (Y  = k \; | \; X = x ) = \hat{p}_k(x) \;\;\;\;\; \forall k \in \{1,...,K\} \;  ,$$

where $(X,Y)$ is a datapoint and $\hat{p} : \mathscr{X} \rightarrow \Delta^K$ returns a probability vector as its output.

This interpretation of calibration aligns the model's prediction with human uncertainty, which means each prediction made by the model is individually reliable and matches human-level uncertainty for that instance. Let's have a look at an example below:

% \clearpage

\begin{figure}[ht]
\begin{center}
    \includegraphics[width=0.9\textwidth]{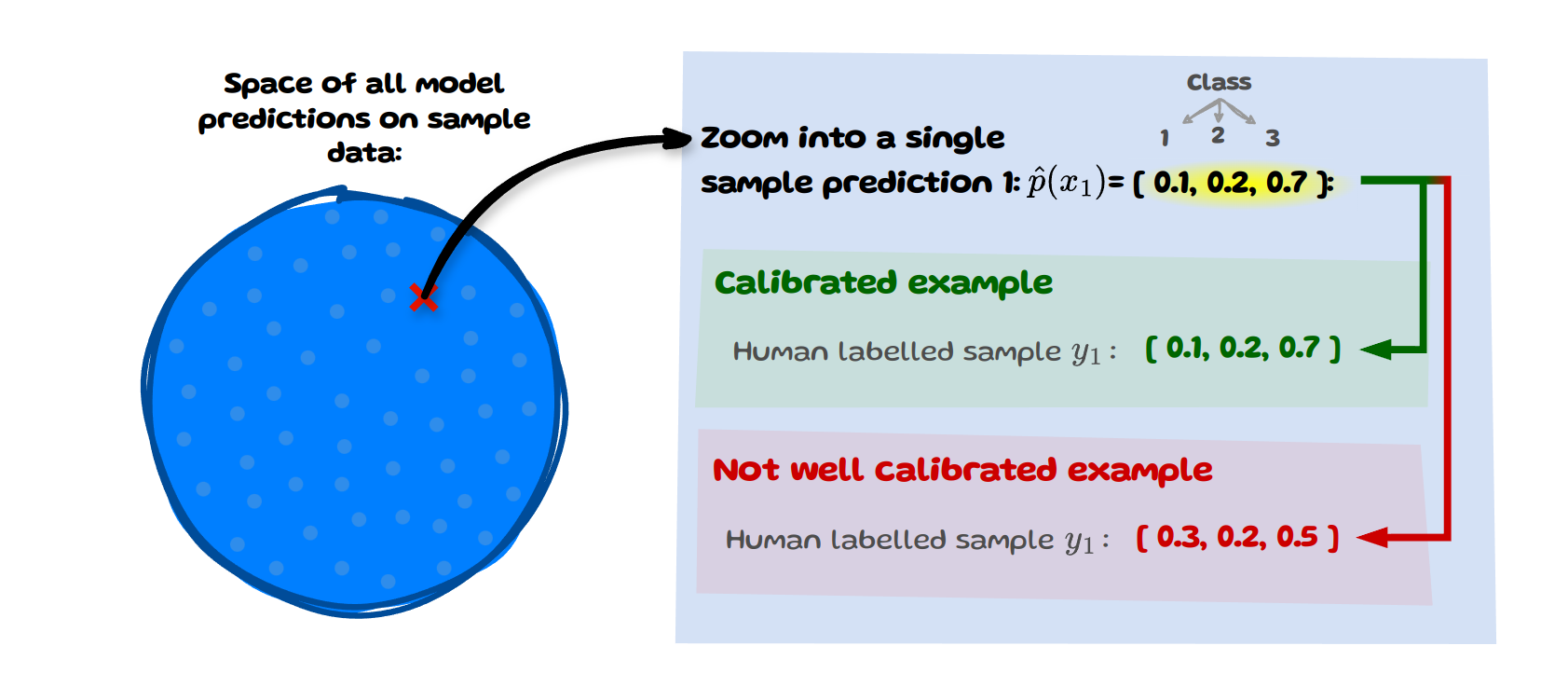}
\end{center}
\caption{Human Uncertainty Calibration}
\label{fig:16}
\end{figure}

We have our sample data (\textit{left}) and zoom into a single sample $x$ with index $i=1$. 
The model's predicted probability vector for this sample is [0.1,0.2,0.7]. 
If the human labelled distribution $y_i$ matches this predicted vector then this sample is considered calibrated.

This definition of calibration is more granular and strict than the previous ones as it applies 
directly at the level of individual predictions rather than being averaged or assessed over a set of samples. It also relies heavily on having  an accurate estimate of the human judgement distribution, which requires a large number of annotations per item. Datasets with such properties of annotations are gradually becoming more available \citep{aroyo2024dices, nie2020learn}.

To evaluate human uncertainty calibration three new measures are introduced by \citet{baan2022stop}: \textbf{the Human Entropy Calibration Error (\textit{EntCE}), the Human Ranking Calibration Score (\textit{{RankCS}) and the Human Distribution Calibration Error (\textit{DistCE})}}.

$$EntCE(x_i)= H(y_i) - H(\hat{p}_i), $$

where $H(.)$ signifies entropy.

\textbf{EntCE} aims to capture the agreement between the model's uncertainty $H(\hat{p}_i)$ and the human uncertainty $H(y_i)$ for a sample $i$. However, entropy is invariant to the permutations of the probability values; in other words it doesn't change when you rearrange the probability values. This is visualised in the image below:
\begin{figure}[ht]
\begin{center}
    \includegraphics[width=0.9\textwidth]{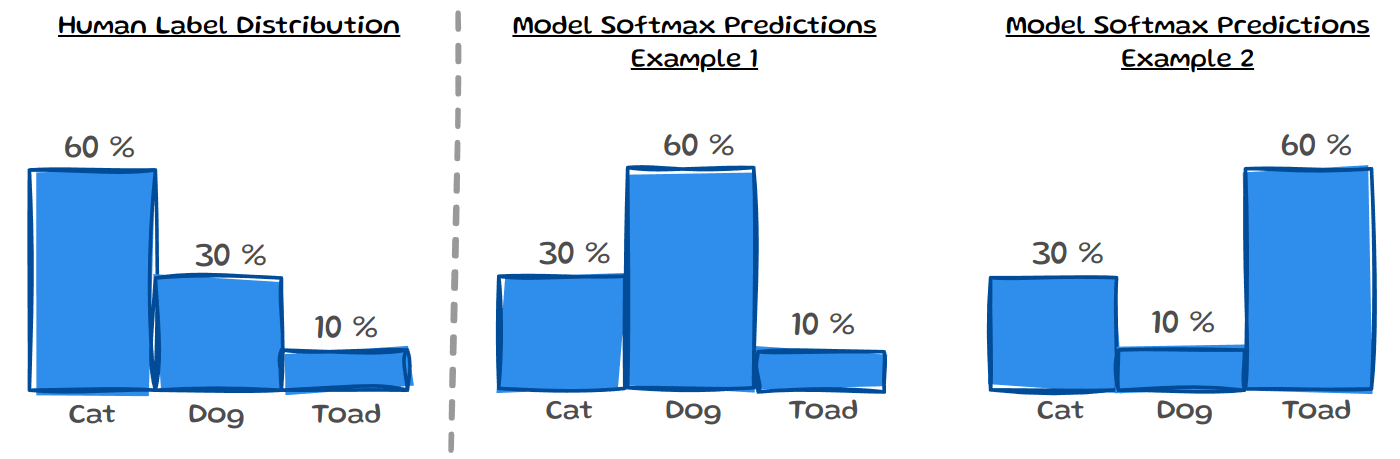}
\end{center}
\caption{EntCE drawbacks}
\label{fig:17}
\end{figure}

On the left, we can see the human label distribution $y_i$, on the right are two different model predictions for that same sample. All three distributions would have the same entropy, so comparing them would result in 0 EntCE. While this is not ideal for comparing distributions, entropy is still helpful in assessing the noise level of label distributions.

$$RankCS = \frac{1}{N} \sum_{n=1}^{N} \mathbf{1} (argsort(y_i) = argsort(\hat{p}_i)), $$   

where argsort simply returns the indices that would sort an array.

So, \textbf{RankCS}  checks if the sorted order of estimated probabilities $\hat{p}_i$ matches the sorted order of $y_i$ for each sample. 
If they match for a particular sample $i$ one can count it as 1; if not, it can be counted as 0, which is then used to average over all samples N. \footnote{In the paper it is stated more generally: If the argsorts match, it means the ranking is aligned, contributing to the overall RankCS score.}

Since this approach uses ranking it doesn't care about the actual size of the probability values. The two predictions below, while not the same in class probabilies would have the same ranking. This is helpful in assessing the overall ranking capability of models and looks beyond just the maximum confidence. At the same time though, it doesn't fully capture human uncertainty calibration as it ignores the actual probability values.

\begin{figure}[ht]
\begin{center}
    \includegraphics[width=0.9\textwidth]{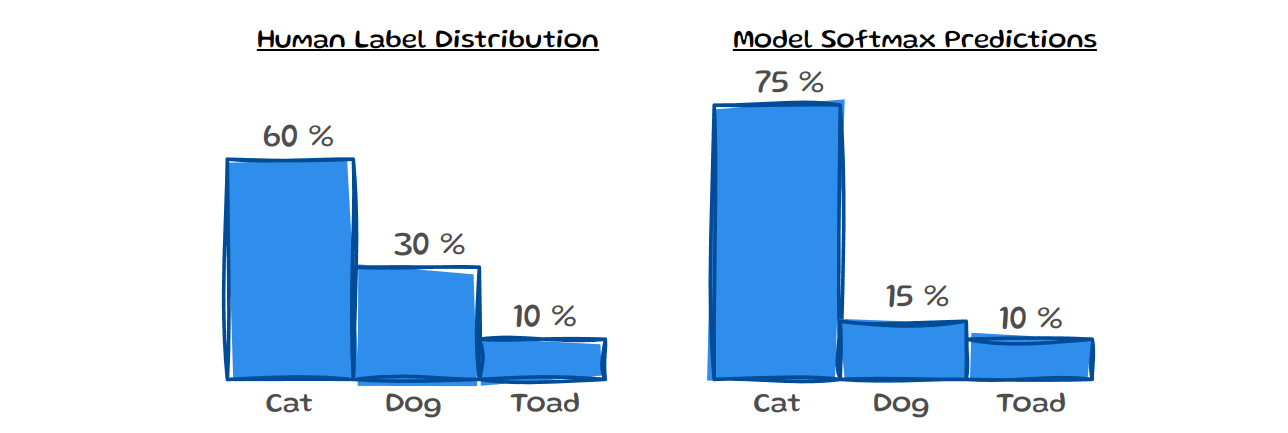}
\end{center}
\caption{RankCS drawbacks}
\label{fig:18}
\end{figure}

\clearpage
% \vspace{0.5cm}

$$ DistCE(x_i) = \mathbf{TVD}(y_i, \hat{p}_i)$$

\vspace{0.5cm}

\textbf{DistCE} has been proposed as an additional evaluation for this notion of calibration. It simply uses the total variation distance $(TVD)$ between the two distributions, which aims to reflect how much they diverge from one another. \textit{DistCE} and \textit{EntCE}  capture instance level information. So to get a feeling for the full dataset one can simply take the average expected value over the absolute value of each measure: $E[\mid DistCE \mid]$ and $E[\mid EntCE \mid]$. 
Perhaps future efforts will introduce further measures that combine the benefits of ranking and noise estimation for this notion of calibration.

\vspace{1cm}
\section{Final Thoughts}

We have run through the most common definition of calibration, the shortcomings of ECE and additional notions of calibration: multi-class, class-wise \& human-uncertainty calibration. 
We also touched on some of the newly proposed evaluation measures and their shortcomings.
Despite several works arguing against the use of ECE for evaluating calibration, it remains widely used. 
The aim of this blogpost is to draw attention to these works and their alternative approaches. 
Determining which notion of calibration best fits a specific context and how to evaluate it should avoid misleading results. 
Maybe, however, ECE is simply so easy, intuitive and just good enough for most applications that it is here to stay?

% \vspace{-1cm}
\begin{figure}[h!]
\begin{center}
    \includegraphics[width=0.94\textwidth]{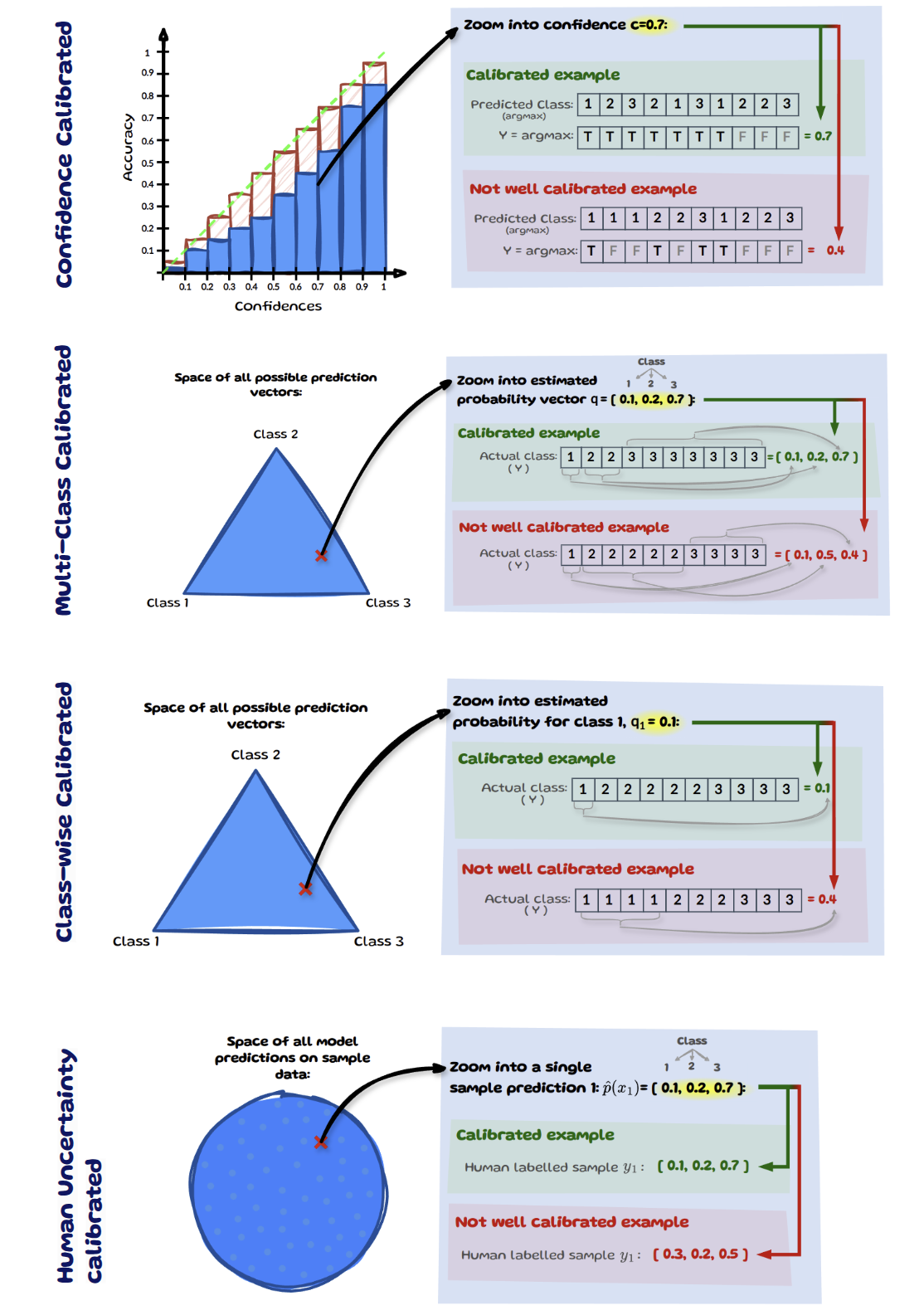}
\end{center}
\caption{Visual Summary: Definitions at a Glance}
\label{fig:19}
\end{figure}

\clearpage

% \vspace{.6cm}
% \noindent\makebox[\linewidth]{\rule{\textwidth}{0.2pt}}
% \vspace{0.1cm}

\textbf{For attribution in academic contexts, please cite this work as}
\lstset{
  language=Reference1,
  basicstyle=\ttfamily\small,
  breaklines=true,
  frame=single,
  columns=fullflexible
}

\begin{lstlisting}
  Pavlovic, "Understanding Model Calibration - A gentle introduction and visual exploration of calibration and the expected calibration error (ECE)", ICLR Blogposts, 2025.
\end{lstlisting}

\textbf{BibTeX citation:}

\vspace{0.1cm}

\lstset{
  language=BibTeX,
  basicstyle=\ttfamily\small,
  breaklines=true,
  frame=single,
  columns=fullflexible
}
\begin{lstlisting}
@inproceedings{pavlovic2025calibration,
  author    = {Pavlovic, Maja},
  title     = {Understanding Model Calibration - A gentle introduction and visual exploration of calibration and the expected calibration error (ECE)},
  booktitle = {ICLR Blogposts 2025},
  year      = {2025},
  date      = {April 28, 2025},
  note      = {https://iclr-blogposts.github.io/2025/blog/calibration/},
  url       = {https://iclr-blogposts.github.io/2025/blog/calibration/}
}
\end{lstlisting}

\vspace{1.1cm}

\bibliography{iclr2025_conference}
\bibliographystyle{iclr2025_conference}

\end{document}